# Intelligent Methods for Accurately Detecting Phishing Websites

Almaha Abuzuraiq
*Computer Science Department*
*Princess Sumaya University for Technology*
Amman, Jordan
alm20178050@std.psut.edu.jo

Mouhammd Alkasassbeh
*Computer Science Department*
*Princess Sumaya University for Technology*
Amman, Jordan
m.alkasassbeh@psut.edu.jo

Mohammad Almseidin
*Computer Science Department*
*University of Miskolc*
Miskolc, Hungary
alsaudi@iit.uni-miskolc.hu

*Abstract*—With increasing technology developments, there is a massive number of websites with varying purposes. But a particular type exists within this large collection, the so-called phishing sites which aim to deceive their users. The main challenge in detecting phishing websites is discovering the techniques that have been used. Where phishers are continually improving their strategies and creating web pages that can protect themselves against many forms of detection methods. Therefore, it is very necessary to develop reliable, active and contemporary methods of phishing detection to combat the adaptive techniques used by phishers. In this paper, different phishing detection approaches are reviewed by classifying them into three main groups. Then, the proposed model is presented in two stages. In the first stage, different machine learning algorithms are applied to validate the chosen dataset and applying features selection methods on it. Thus, the best accuracy was achieved by utilizing only 20 features out of 48 features combined with Random Forest is 98.11%. While in the second stage, the same dataset is applied to various fuzzy logic algorithms. As well the experimental results from the application of Fuzzy logic algorithms were incredible. Where in applying the FURIA algorithm with only five features the accuracy rate was 99.98%. Finally, comparison and discussion of the results between applying machine learning algorithms and fuzzy logic algorithms is done. Where the performance of using fuzzy logic algorithms exceeds the use of machine learning algorithms.

*Keywords*—Phishing detection, phishing website, machine learning, fuzzy logic, feature selection.

## I. INTRODUCTION

The Internet is everywhere today, everyone uses web services for a range of activities such as sharing knowledge, social communication and performing various financial activities which include buying, selling and money transferring. Malicious websites are a serious threat to internet users and unaware users can become victims of malicious URLs that host undesirable content such as spam, phishing, drive-bydownload, and drive-by-exploits.

Phishing websites are the most common malicious websites that attend to luring users into fraudulent attempts to obtain their personal or sensitive information. The number of these websites is continuously increasing over time. Where according to the Anti-Phishing Working Group (APWG) report, the number of different phishing incidents reported to the organization over the last quarter of the year 2016 was 211,032 [1]. Additionally, they were increased by 12% in the last quarter of 2018 which received 239,910 reports [2]. Phishing websites are a serious problem against individuals and corporations. Where harmful consequences could result. For individuals, the attacker can use their information in unauthorized purchases, funds stealing or theft of identity. While, for corporations, the employees are compromised to bypass the security circumferences, spread malware inside a corporate environment, access secured data. When a phishing attack occurred on the organization's system their market share will get lower, as well their reputation and customer's trust will be affected [3].

The need for detecting these kinds of websites and alert users to protect themselves is very necessary. Therefore, this paper concentrates on the problem of detecting if the websites are phishing or legitimate by utilizing different methods. The major challenge in detecting phishing attacks is discovering the utilized techniques. Accordingly, developing robust, effective and contemporary phishing detection methods are very necessary to oppose the adaptive techniques employed by the phishers [4].

By surveying the previous studies of phishing detection techniques, they are categorized into the following approaches: Blacklist based, Content-based, Heuristic-based, and Fuzzy rule-based approaches. Some of the existing phishing detection approaches are suffering from low detection accuracy particularly when using the blacklisted approach which is inefficient in responding to novel phishing attacks. Where producing a new domain now is much easier. Moreover, checking the page contents has been used to solve the problem of the blacklistedbased approach in general, but also it is not sufficient; because one of their drawbacks is the dependency on third-party servers and can't work as a standalone approach. Also, it assaults the user's privacy by uncovering his browsing history [4]. As well, the heuristic-based approach has a limitation that It takes a very long time to extract and develop a large number of features. Each approach has its advantages and disadvantages. Accordingly, It is very important to continue improving phishing detection methods in order to get as possible as an accurate, fast and efficient approach. In this paper, a fuzzy logic-based approach will be utilized and compared with a heuristic-based approach using machine learning algorithms.

The rest of the paper is organized as follows: section (II) illustrates the background of the concept of phishing. The related works of phishing detection approaches are demonstrated in section (III). Followed by the details of the proposed method in a section (IV). After that, section (V) shows the evaluation and discussion of the experimental results, and finally the conclusion in section (VI).

## II. BACKGROUND

Phishing is defined as a form of social engineering attacks in which an attacker attempt to steal the user's data like username, password, social security number or credit card number. The attacker pretends to be a legitimate entity, a specific person in email or other communication methods in order to trick the victim to open an email, instant or text messages. The victim is tricked to click a malicious link that may install the malware on the user's device, freezing the system or even uncover the victim's sensitive data [3]. For example, a phishing email that appears to come from an authorized bank to alert the recipient that their account information has been exposed, leading the victim to a fraudulent website that designed to look legitimate to reset their username and/ or password. This website is maintained to gather login information from victims of phishing. Some of the most common forms of phishing attacks are:

- Email phishing: is one of the easiest phishing types that used to obtain the user's data. There are various of methods for Email phishing such as, sending an email from a known username, sending an email pretending to be the manager and asking for a significant data or impersonating an organization and asking their employees to share private data [5].
- Spear phishing: is a targeted phishing form, unlike random phishing emails, it is directed at a specific person or company. The attackers use social engineering techniques to personalize the emails according to their victims. In addition, it is a very important form of phishing, and 91% of cyber attacks started with spear-phishing emails
  [6].
- Whaling: is a form of spear phishing that is more specific, where the attacker addressed it to the senior executives within an organization. For example, targeting the employees with the capability to authorize payments, by a phishing message that appears as an order from the executive authorize a massive payment to a customer, but the payment would be sent to the attackers [7].
- Clone phishing: The clone phishing attack takes benefits of the legitimate messages that already received by the victim and produce a malicious copy of it, then send the message from an email that appears to be a legitimate one. All URLs or attached files in the native message are replaced with malicious ones. After that, the attackers tell the victim that they resend the original message because there was a problem with the attachments to deceive the victim to reopen them [8].

## III. RELATED WORKS

Many researchers proposed different phishing detection techniques. In the following section, some of the existing approaches will be reviewed which are a part of our previous study in [9]. where have been categorized into three groups, which are: Content-Based Approach, Heuristic Based Approach, and Fuzzy rule-based approach.

### A. Content-Based Approach

This approach works on a deep analysis of the pages' content. Building classifiers and extract features from page contents and third-party services such as search engines, and DNS servers. The traditional anti-phishing methods are based on visual similarities which are effective only in detecting phishing web pages that show a high similarity rate in their contents with the counterpart legitimate web page.

Therefore, the authors in [10] proposed an unprecedented method for detecting phishing web pages by specifying weights to the words that draw out from URLs and HTML contents. These words may include brand names that phishers attempt to set them in various parts of the URL to make it looks like the real one. Weights are specified according to their presence at different positions in URLs. Then, these weights are combined with their term frequency-inverse document frequency (TF-IDF) weights, which is a numeric statistic that shows how significant a word is to a document. The most probable words are chosen and send to Yahoo Search to return the domain name that has the highest frequency between the top 30 outcomes. Eventually, they decide if the website is authentic or not by comparing the owners of the domain name that returned from WHOIS records. A WHOIS lookup is applied to detect the owner of such a domain name.

Instead of utilizing a brand name and word weights, the paper of [11] presented an unprecedented approach that utilizes a logo image to determine the identity of the web page by matching real and fake webpages. Where the proposed approach is composed of two phases which are logo extraction and identity verification. In the first phase, the machine learning algorithm will extract the logo image from all the downloaded image resources of the website. Then in the second phase by using google image search it will retrieve the identity of the extracted logo image, and because of the exclusive relationship between the logo and domain name, they treat the domain name as the identity. Hence, a comparison between the domain name returned by Google with the one from the query website. The system evaluated by using two different datasets is made of 1140 phishing sites obtained from Phish Tank and legitimate websites obtained from Alexa. The accuracy of the proposed system is 93.4%.

## B. Heuristic-Based Approach

This detection approach is based on employing various descriptive features extracted by understanding and analyzing the structure of phishing web sites. The method used in processing these features plays a considerable role in classifying web sites effectively and accurately [12].

Due to the importance of defining a valuable and clear feature, the paper of [13] proposed a new model to detect phishing websites using six heuristics features extracted from URLs (primary domain, subdomain, path domain) and website rank (page rank, Alexa rank, Alexa reputation). Also, this approach is evaluated by utilizing a training dataset of 11,660 phishing web sites and ten testing datasets, each one holds 1,000 phishing web sites or 1,000 legitimate web sites. The experiment results exhibit that the accuracy of the proposed system in detecting phishing web sites is 97.16%.

Therefore, in [14] the author demonstrates hybrid machine learning approaches that get a benefit from the strengthens of each algorithm and neglect its weaknesses. These algorithms are K-nearest neighbors (KNN) algorithm which is an effective approach against unwanted data, and the Support Vector Machine (SVM) algorithm which is a robust classifier. The combination process is done in two phases. At first, KNN will be applying, then SVM will employing as a classification tool. The dataset used for the experiment is taken from a related work that contains more than 1353 samples gathered from various sources. Each sample record is composed of nine features in addition to the class label which is Phishing, Legitimate, or Suspicious web sites. Consequently, the clearness of KNN is integrated with the effectiveness of SVM, regardless of their disadvantages when they used individually. The accuracy of the proposed method is 90.04%.

## C. Fuzzy Rule-Based Approach

The Fuzzy logic permits the intermediate level among values, and it is utilized to classify web pages based on the level of phishing, that appeared in the pages by employing a specific group of metrics and predefined rules [15]. Using this approach allows the processing of ambiguous variables, then integrates human experts to clarify those variables and relations between them. Also, fuzzy logic approaches using linguistic variables to explain phishing features and the likelihood of phishing web page [16].

In trying to get the benefits of a fuzzy logic system, the paper of [17] proposed a novel approach that targeted the URL features and fuzzy logic method. The system is applied in five phases which are: select URL features, calculating the values of 6 heuristics, calculating 12 fuzzy values for 6 heuristics from membership functions, defuzzification by calculating mean of 6 fuzzy values of phishing linguistic label (MP) and mean of 6 fuzzy values of legitimate linguistic label (ML). Finally, the values of MP and ML will be compared to classify the web page. The approach was assessed with 11,660 phishing web pages and 5,000 legitimate web pages. The accuracy of the proposed method was 98.17%.

Instead of using a standalone fuzzy system, in [18] the authors applied a Neuro-Fuzzy Scheme, which is an integration of a Fuzzy Logic and a neural network Unless a data set of 300 value is extracted from six data sources which are Legitimate site rule, User behavior profile, PhishTank, User-specific-site, Pop-up windows, and User-credential profile. There are the same as the previous study, but a new source was added, which is the User-credential profile. In addition, the proposed system applied using 2-fold cross-validation to training and testing the model. The fuzzy model has five functions to understand and judge which include, input layer, fuzzification, rule-based, normalization, and defuzzification. The proposed system achieved 99.6% of accuracy which is better than the previous study. It can be concluded that even the web page contains phishy characteristics it does not mean that the whole page is phishy. Therefore, using fuzzy logic is one of the most effective methods to obtain the phishiness degree of a web page. Table I gives a summary of some of the related works.

TABLE I SUMMARY OF THE RELATED WORKS.

| Paper | Approach | Drawback | Accuracy |
|---|---|---|---|
| [10] | Specifying weights to the words that draw out from URLs and HTML contents such as Brand name | Dependency on third party server which is Yahoo Search. | 98.20% |
| [11] | Utilizing a logo image to determine the identity of the web page by matching real and fake webpages | Dependency on third party server which is google image search | 93.40% |
| [13] | Using URLs heuristics and website rank | Taking long time in extracting the features and look after website rank. | 97.16% |
| [14] | Proposed a method that combines two algorithms. *KNN* and *SVM*. | Although it takes high computation time, the accuracy rate was low according to similar studies | 90.04% |
| [17] | Using of a fuzzy logic system | The system is applied in five phases. Which may take a considerable time to be built. | 98.17% |
| [18] | Employing fuzzy systems and neural networks | Setting rules and membership functions are a challenging task | 99.6% |

## IV. THE PROPOSED METHOD

In building phishing detection systems, there are two main important parts which are the algorithms utilized to build the model and the data set employed for training and testing it. Although the main purpose of this study is building a phishing detection system using a fuzzy logic algorithm. But the dataset chosen to train and test the system has not been used by many researchers yet. Therefore, the validity of the dataset will be ensured by testing it on various machine learning algorithms. Then, different features selection methods will be applied on that dataset in order to enhance the model's performance. After that different four fuzzy logic algorithms will be applied on the same data set. Finally, the experimental results from both approaches will be compared and discussed.

*A. Data set*

The data set used in this paper is offered by [19] is composed of 5000 phishing websites and 5000 legitimate websites. Phishing websites are collected from PhishTank and OpenPhish, while legitimate websites are collected from Alexa and Common Crawl. As well the data set consists of 48 features that were extracted by utilizing the website's URLs and HTML source codes.

*B. Feature selection*

Feature selection is one of the basic machine learning principles where the features used to train the models have a significant impact on the efficiency of the system. The main advantage of using feature selection methods is that it prevents overfitting, and helps the model to concentrate only on the important features by eliminating unnecessary ones. In this paper, two algorithms have been applied to the dataset which are Infogain and Relief-F. Both of them are of the filter method's type. Where the most important benefits of the filter methods that they have a very low computation time and do not overfit the data.

In utilizing the two algorithms, the top 15 features resulted from each algorithm had been considered. Then the top 5 and top 10 were extracted from them to apply different experiments on the machine learning algorithms and observing the accuracy of each. Finally, in order to get better accuracy rates, the experiments were also applied to the union and the intersection of the top 15 features. Where 20 features resulted from the union phase and 10 features resulted from the intersection.

*C. Model evaluation*

To evaluate the models there are many assessment tools. But in this paper, the model evaluated by using the accuracy equation because the utilized dataset is a binary and balanced data set. Thus, calculating the accuracy rate only is enough. Accuracy is calculated using the following equation (1).

$$Accuracy = \frac{\sum truepositive + \sum truenegative}{\sum totalpopulation} \quad (1)$$

*D. Machine Learning*

Initially, machine learning has defined as the computer's ability to learn and make a decision. Arthur Samuel in 1959 defined machine learning as the "field of study that gives computers the ability to learn without being explicitly programmed". Machine learning is a study that combines [20].

To achieve the main purpose of implementing machine learning algorithms in this study is to evaluate the dataset, various experiments on different machine learning classifiers were performed. The used algorithms are Bayes net, Naïve Bayes, J48, Logistic, Random forest, Bagging, and Multilayer perceptron. However, only three algorithms have been chosen, that obtained the best accuracy rates. As well, based on the previous studies these algorithms are the most commonly used for the purpose of phishing detection, which are J48, Random forest(RF), and Multilayer perceptron(MLP).

*E. Machine learning experimental results*

To run the experiments a machine with a specification Intel(R) Core (TM) i5-2450M CPU @2.50GHz (2 processors) has been used. In training and testing the machine learning algorithms the whole dataset and the datasets resulted from different feature selection processes have been applied. As well, for analyzing and comparing the classifiers, a data mining tool called Weka 3.8.3 has been utilized. The results of using the three different algorithms with nine different datasets are shown in figure 1. Note that the details about machine learning experimental results are shown in the paper of [21]. By

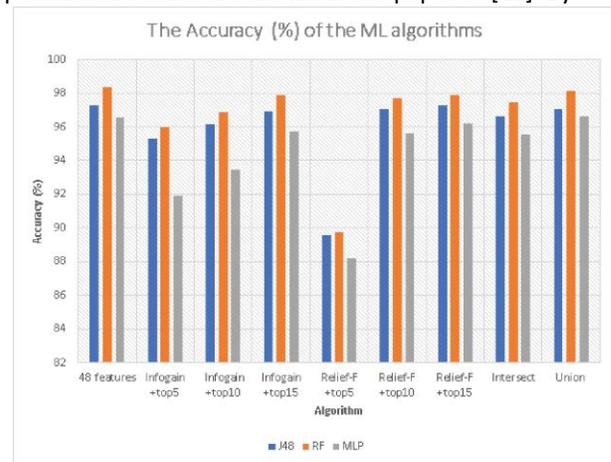

Fig. 1. The accuracy of the ML algorithms. comparing the results of using the three different algorithms with nine different datasets. The findings can be concluded on some points as follow:

- The higher the number of features the higher accuracy rates.
- The order of the best algorithm according to the best average accuracy rates is RF, J48, then MLP.
- The order of the algorithms in accordance to the less time taken on building the model are J48, RF, then MLP.
- Applying all the features to the RF algorithm was the highest accuracy of all 27 experiments. But the best accuracy resulted from applying feature selection methods was by applying also RF to the Union features. Where applying the whole features showed 98.37% and applying the Union features showed 98.11%.

*F. Fuzzy Logic*

Fuzzy logic is based on the theory that the absolute "black and white" doesn't exist when calculating probabilities. And it is a sort of infinite-valued logic that deals with uncertain rather than exact values. In addition, fuzzy logic was introduced by Zadeh in the late 1960s as a reinvention of Lukasiewicz's multi-valued logic which is a mathematical calculation in which there are more than two truth values (true or false) [22]–[24].

To apply the fuzzy logic system, the inputs must convert to fuzzy inputs. Then, applying the rules of the system. After that, combining the results obtained from different rules using the aggregation function. Finally, using the inference function to defuzzificate the aggregated results. Accordingly, these processes of applying a fuzzy logic system can be classified by four main sections:

1) Initialization process:
    a) Specifying the linguistic variables.
    b) Create the functions of fuzzy logic membership that determine the meaning of the terms used in the rules for input and output.
    c) Build the rule base which constructs IF-THEN rules.
2) Use the membership functions to transform crisp input data into fuzzy values.
3) Assess the rules in the rule base and collect the results of each rule by finding the corresponding degree with each rule between the current fuzzy input.
4) Convert the output data to values that are crisp, this process is known as defuzzification. [25]–[27]. See figure 2.

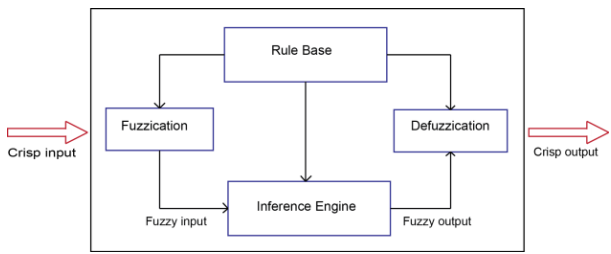

Fig. 2. The architecture of the fuzzy logic system

Four fuzzy logic algorithms were utilized to find the accurate and efficient fuzzy logic algorithm in phishing detecting purposes. Two of them are evolutionary fuzzy rule learning algorithms, which are Hybrid fuzzy GBML and Adaboost algorithm. In addition to the two standard fuzzy rule learning algorithms which are Chi-rule weighted algorithm and FURIA. These algorithms were chosen according to the most commonly used in previous studies.

*G. Fuzzy logic experimental results*

For the purpose of assessing the fuzzy logic algorithms, the experiments have been applied using KEEL software. As well the machine specification used in running the experiments is Intel (R) Xeon (R) CPU E5-2680 @ 2.70 GHz (2 processors). the results of using the four different algorithms with only seven datasets which are top 5 and top 10 of using Infogain and Relief-F, intersection and union are shown in figure 3.

By comparing the experimental results of using the different fuzzy logic algorithms, the observations can be concluded in some points:

- The higher the number of features, the less accuracy. Where adapting the fuzzy model as a decision system depends on many factors such as the type of inference system and the required input parameters. Each input parameter in the fuzzy model has a weight which will affect directly the results of the fuzzy system. Therefore, using a large number of features which include a number of irrelevant features will decrease the performance of fuzzy model.
- The classifier with the lowest average of the accuracy rates is AdaBoost. Even though it gave 100% accuracy

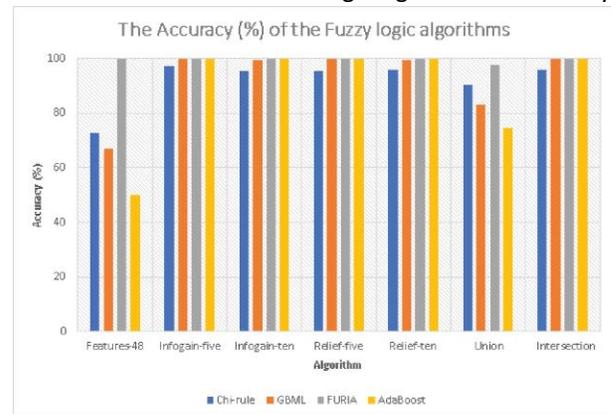

Fig. 3. The architecture of the fuzzy logic system

rates when using only five features in both Infogain and Relief-F. And almost 100% in the rest of the datasets except using the union of Infogain and Relief-F and the whole dataset. But the remaining accuracy rates were lower and very various.
- As aforementioned in the first point, by using the total number of features, the performance of the model could be decreased. The main disadvantage of this algorithm that is it could be sensitive to noisy data and outliers. Therefore, selecting the suitable features selection algorithm is very important to adapt to the Adaboost algorithm.
- The FURIA is the classifier with the highest accuracy rate. Where it achieved an accuracy rate of 99.98%. The results

of adapting FURIA were very close which means that it could not be sensitive to the outliers on the contrary with the AdaBoost algorithm.
- The conclusion, in implementing fuzzy logic systems, all features will generate their membership functions and related rules. Thus, if the employed fuzzy logic algorithm is affected by irrelevant features, the effectiveness of the classifier will be decreased and vice versa.

V. EVALUATION AND DISCUSSION

According to the observations taken from the experiments. The main outcomes from applying machine learning algorithms and the fuzzy logic algorithms will be shown below:
- Unlike machine learning algorithms where the higher number of features, the higher accuracy rate. In fuzzy logic systems, the lower number of features leads to a higher accuracy rate.
- The best accuracy rate by using machine learning algorithms was when employing the Random forest algorithms with the Union of the top fifteen features using Relief-F and Infogain which results twenty features. Regardless, the result of using the Union criteria was very close to utilizing the whole data set, but the time taken to build the classifier was shorter, almost.
- The average accuracy rates in each algorithm seem to be close in using machine learning algorithms. While in using the fuzzy logic algorithms, the average rates are very varied. The reason that some fuzzy logic algorithms is highly affected by the unrelated features.
- The experimental results gained from applying fuzzy logic algorithms were very incredible. Where the accuracy rates were close to 100% in some experiments that have been utilized only five features. This means that the time taken in building the model was lower, in contrast with employing twenty features as in the case of applying the Random forest algorithm.

VI. CONCLUSION

There are many existing websites with a variable purpose, one of these websites is called phishing sites. The phishers intent to trick individuals, corporations, or financial organizations. That may cause a massive financial loss besides to inject malware on the device of the victims or freezing their systems. The need for discovering these websites and alert users is very necessary.

In this paper, the different phishing detection approaches have been classified into three groups which are: ContentBased approach, Heuristic-Based approach, and Fuzzy rulebased approach. Then, the proposed methodology was classified into two phases. First, various machine learning algorithms. The best result was achieved by utilizing a Random forest algorithm on the 20 features that resulted from the union's process of the top fifteen features. The obtained accuracy was 98.11%. While in the second phase, the same dataset was applied to different fuzzy logic algorithms. Whereas two of them are evolutionary fuzzy rule learning algorithms. In addition to two standard fuzzy rule learning algorithms. The experimental results of using Fuzzy logic algorithms were unexpectable. Where some experiments had approximately 100 % of accuracy rates by applying only five features. Finally, the results obtained from using both machine learning algorithms and fuzzy logic algorithms were compared and discussed.